\documentclass[11pt,a4paper]{amsart}
\usepackage{hyperref}
\hypersetup{
    pdftitle={Geodesic Flow on the Diffeomorphism Group of the circle}, 
    pdfauthor={A. Constantin and B. Kolev} 
    }
\newtheorem{thm}{Theorem}

\newtheorem{lem}{Lemma}
\newtheorem{prop}{Proposition}
\theoremstyle{definition}

\theoremstyle{remark}

\numberwithin{equation}{section}
\newcommand{\norm}[1]{\left\Vert#1\right\Vert}

\newcommand{\abs}[1]{\left\vert#1\right\vert}
\newcommand{\set}[1]{\left\{#1\right\}}

\newcommand{\Circle}{\mathbb{S}}
\newcommand{\Diff}{\mathcal{D}}

\begin{document}

\title[Geodesic flow]{Geodesic flow on the diffeomorphism group of the circle}

\author[A. Constantin]{Adrian Constantin}
\address{Lund University, Dept. of Maths. P.O  Box 118, S-22100 Lund, Sweden}
\email{adrian.constantin@math.lu.se}

\author[B. Kolev]{Boris Kolev}
\address{CMI, 39, rue F. Joliot-Curie, 13453 Marseille cedex 13, France}
\email{kolev@cmi.univ-mrs.fr}%


\begin{abstract}
We show that certain right-invariant metrics endow the
infinite-dimensional Lie group of all smooth
orientation-preserving diffeomorphisms of the circle with a
Riemannian structure. The study of the Riemannian exponential map
allows us to prove infinite-dimensional counterparts of results
from classical Riemannian geometry: the Riemannian exponential map
is a smooth local diffeomorphism and the length-minimizing
property of the geodesics holds.
\end{abstract}
\maketitle


\section{Introduction}

The group $\Diff$ of all smooth orientation-preserving
diffeomorphisms of the circle $\Circle$ is the ``simplest
possible'' example of an infinite-dimensional Lie group
\cite{ArnKhe}. Its Lie algebra $T_{Id}\Diff$ is the space
$C^\infty(\Circle)$ of the real smooth periodic maps of period
one. Since $C^{\infty}(\Circle)$ is not provided with a natural
inner product, to endow $\Diff$ with a Riemannian structure we
have to define an inner product on each tangent space
$T_{\eta}\Diff,\ \eta \in \Diff$. For a Lie group the Riemannian
exponential map of any two-sided invariant metric coincides with
the Lie group exponential map \cite{ArnKhe}. It turns out that the
Lie group exponential map on $\Diff$ is not locally surjective cf.
\cite{Ham} so that a meaningful\footnote{~A prerequisite of a
rigorous study aimed at
  proving infinite-dimensional counterparts of facts established in
  classical (finite-dimensional) Riemannian geometry is the use of the
  Riemannian exponential map as a local chart on $\Diff$.}
Riemannian structure cannot be provided by a bi-invariant metric
on $\Diff$. We are led to define an inner product on
$C^{\infty}(\Circle)$ and produce a right-invariant
metric\footnote{~To carry out the passage from right-invariant
metrics
  to left-invariant ones, note that a right-invariant metric on
  $\Diff$ is transformed by the inverse group operation to a
  left-invariant metric on $\Diff$ with the reverse law
  ($\varphi \star \psi=\psi \circ \varphi$).} by
transporting this inner product to all tangent spaces
$T_{\eta}\Diff,\ \eta \in \Diff$, by means of right translations.

In this paper we show that certain right-invariant metrics induce
noteworthy\footnote{~Interestingly, for a particular metric in
this
  class, the corresponding geodesic equation is encountered in
  hydrodynamics as a model for the unidirectional propagation of
  shallow water waves - for an elaboration of this viewpoint and its
  implications we refer to \cite{ConKol}.} Riemannian structures on
$\Diff$. Despite the analytical difficulties that are
inherent\footnote{~$\Diff$ is a Fr{\'e}chet manifold so that the
  inverse function theorem and the classical local existence theorem
  for differential equations with smooth right-hand side are not
  granted cf. \cite{Ham}. Moreover, we deal with weak Riemannian
  metrics (the family of open sets of $\Diff$ contains but does
  not coincide with the family of open sets of the topology induced by
  the metric) so that even the existence of a covariant derivative
  associated with the right-invariant metric is in doubt
  cf. \cite{Lan}.}, the existence of geodesics is obtained and their
length-minimizing property is established. The paper is organized
as follows. In Section 2 we discuss the manifold and Lie group
structure of $\Diff$, Section 3 is devoted to basic properties of
the Riemannian structures that we construct on $\Diff$ (existence
and local chart property of the Riemannian exponential map), while
in Section 4 we prove the length-minimizing property of geodesics.
In the last section we present\footnote{~For a detailed analysis
see \cite{ConKol}.} a choice of a right-invariant matric endowing
$\Diff$ with a deficient Riemannian structure (the corresponding
Riemannian exponential map is not a local $C^1$-diffeomorphism),
to emphasize the special features of the previously discussed
Riemannian structures. Note that for diffeomorphism groups the
existence of geodesics is an open question\footnote{~For any
smooth compact manifold $M$, both the group of smooth
diffeomorphisms of $M$, $\hbox{Diff}(M)$, and its subgroup formed
by the volume-preserving diffeomorphisms, have a Lie group
structure \cite{Mil}. Progress towards the existence of geodesics
was made \cite{EbiMar} but the understanding is still incomplete
\cite{ArnKhe}.} so that it is of interest to have an example
($M=\Circle$) where an attractive geometrical structure is
available.


\section{The diffeomorphism group}

In this section we discuss the manifold and Lie group structure of
$\Diff$. If $\xi(x)$ is a tangent vector to the unit circle
$\Circle$ at $x\in \Circle \subset \mathbb{C}$, then $\Re\,
[\overline{x}\ \xi (x) ] = 0$ and
\begin{equation*}
u(x) = \frac{1}{2\pi i}\
\overline{x}\  \xi (x) \in \mathbb{R}.
\end{equation*}
This allows us to identify the space of smooth vector fields on
the circle with $C^\infty(\Circle)$, the space of real smooth maps
of the circle. The latter may be thought of as the space of real
smooth periodic maps of period one and will be used as a model for
the construction of local charts on $\Diff$. Note that
$C^\infty(\Circle)$ is a Fr{\'e}chet space, its topology being defined
by the countable collection of $C^n(\Circle)$-seminorms: a
sequence $u_j \to u$ if and only if for all $n \ge 0$ we have $u_j
\to u$ in $C^n(\Circle)$ as $j \to \infty$. $\Diff$ is an open
subset of $C^\infty(\Circle;\Circle)\subset
C^\infty(\Circle;\mathbb{C})$, as one can easily see considering
the function defined on $C^\infty(\Circle;\Circle)$ by
\begin{equation*}
\Theta (\varphi) = \inf_{x\neq y} \frac{\mid \varphi (x) - \varphi
(y)\mid}{\mid x-y \mid}.
\end{equation*}
We will describe a Fr{\'e}chet manifold structure on $\Diff$. If
$t\mapsto \varphi (t)$ is a $C^1$-path in $\Diff$ with $\varphi
(0) = Id$, we have $\varphi'(0)(x) \in T_x \Circle$. Therefore
$\varphi' (0)$ is a vector field on $\Circle$ and we can identify
$T_{Id} \Diff$ with $C^\infty(\Circle)$. If $\varphi \in \Diff$ is
such that $\Vert \varphi-Id \Vert_{C^0(\Circle)}<1/2$, we can
define
\begin{equation*}
u(x)=\frac{1}{2\pi i}\ \ln\Bigl(\overline{x}\  \varphi
(x)\Bigr)\in C^{\infty}(\Circle;\mathbb{R}).
\end{equation*}
Note that $u(x)$ is a measure of the angle between $x$ and
$\varphi(x)$. Choose a lift $F:\mathbb{R} \to \mathbb{R}$ of
$\varphi$ such that
\begin{equation*}
u\circ \Pi (\tilde{x})=F(\tilde{x})-\tilde{x},\quad \tilde{x} \in
\mathbb{R},
\end{equation*}
where $\Pi:\mathbb{R} \to \Circle$ is the cover map. In the
neighborhood
\begin{equation*}
U_0=\set{\varphi \in \Diff; \quad \norm{
\varphi-\varphi_0}_{C^0(\Circle)}<1/2}
\end{equation*}
of $\varphi_0 \in \Diff$ we are led to define
\begin{align*}
u(x)=\frac{1}{2\pi i} \ln \Bigl(\overline{\varphi_0(x)} \varphi
(x)\Bigr),\quad x \in \Circle,
    \intertext{and} u \circ \Pi
(\tilde{x})=F(\tilde{x})-F_0(\tilde{x}),\quad \tilde{x} \in
\mathbb{R}.
\end{align*}
For $\varphi \in U_0$, let $\Psi_0(\varphi)=u$. We obtain the
local charts $\{ U_0,\Psi_0 \}$, with the change of charts given
by\footnote{~With $u_1=\Psi_1(\varphi)$ and $u_2=\Psi_2(\varphi)$
for $\varphi \in U_1 \cap U_2$, we have
\begin{multline*}
(2 \pi i )\, u_2=\ln(\overline{\varphi_2}\varphi)
=\ln(\overline{\varphi_2}\varphi_1
\overline{\varphi_1}\varphi)=\ln(
\overline{\varphi_2}\varphi_1)+\ln
(\overline{\varphi_1}\varphi)=\ln(
\overline{\varphi_2}\varphi_1)+(2 \pi i )\, u_1.
\end{multline*}
Hence $\Psi_2(\varphi)=\Psi_1(\varphi) + \frac{1}{2\pi i}\
\ln(\overline{\varphi_2}\ \varphi_1)$ and the change of charts is
plain.}
\begin{equation*}
\Psi_2 \circ \Psi_1^{-1}(u_1)=u_1+\frac{1}{2\pi i}\
\ln(\overline{\varphi_2}\ \varphi_1).
\end{equation*}
The previous transformation being just a translation on the vector
space $C^\infty(\Circle)$, the structure described above endows
$\Diff$ with a smooth manifold structure based on the Fr{\'e}chet
space $C^\infty(\Circle)$.

A direct computation shows that the composition and the inverse
are both smooth maps from $\Diff\times \Diff\rightarrow \Diff$,
respectively $\Diff\rightarrow\Diff$, so that the group $\Diff$ is
a Lie group. Note that the derivatives at the Identity of the
left-translation
\begin{align*}
L_{\eta}: \Diff \to \Diff,\quad L_{\eta}(\varphi)=\eta \circ
\varphi,\qquad \eta\in \Diff,
    \intertext{and right-translation}
R_{\eta}: \Diff \to \Diff,\quad R_{\eta}(\varphi)=\varphi \circ
\eta,\qquad \eta \in \Diff,
    \intertext{are given by}
L_{\eta^*}:T_{Id}\Diff\rightarrow T_{\eta}\Diff,\quad u \mapsto
\eta_x\cdot u,\qquad \eta \in \Diff,
    \intertext{respectively}
R_{\eta^*}:T_{Id}\Diff\rightarrow T_{\eta}\Diff,\quad u \mapsto u
\circ \eta,\qquad \eta \in \Diff.
\end{align*}

The Lie bracket on the Lie algebra $T_{Id}{\Diff} \equiv
C^\infty(\Circle)$ of $\Diff$ is
\begin{equation}\label{LieBracket}
[u,v]=-(u_xv-uv_x),\qquad u,v \in C^\infty(\Circle).
\end{equation}
Each $v \in T_{Id}\Diff$ gives rise to a one-parameter group of
diffeomorphisms $\{ \eta(t,\cdot)\}$ obtained solving
\begin{equation}\label{OneParameterSubGroup}
\eta_t=v(\eta)\quad\hbox{in}\quad C^\infty(\Circle)
\end{equation}
with data $\eta(0)=Id \in \Diff$. Conversely, each one-parameter
subgroup $t \mapsto \eta(t) \in \Diff$ is determined by its
infinitesimal generator $v=\frac{\partial}{\partial t}\, \eta(t)
\Bigl|_{t=0} \in T_{Id}\Diff$. Evaluating the flow $t \mapsto
\eta(t,\cdot)$ of \eqref{OneParameterSubGroup} at $t=1$ we obtain
an element $\exp_L(v)$ of $\Diff$. The Lie-group exponential map
$v \to \exp_L(v)$ is a smooth map of the Lie algebra to the Lie
group. Although the derivative of $\exp_L$ at $0 \in
C^\infty(\Circle)$ is the identity, $\exp_L$ is not locally
surjective cf. \cite{Mil}. This failure, in contrast with the
finite-dimensional case, is due to the fact that the inverse
function theorem does not necessarily hold in Fr{\'e}chet spaces cf.
\cite{Ham}.

Let $\mathcal{F}(\Diff)$ be the ring of smooth real-valued
functions defined on $\Diff$ and $\mathcal{X}(\Diff)$ be the
$\mathcal{F}(\Diff)$-module of smooth vector fields on $\Diff$.
For $X \in \mathcal{X}(\Diff)$ and $f \in \mathcal{F}(\Diff)$, the
Lie derivative $\mathcal{L}_Xf$ is defined in a local chart as
\begin{equation*}
\mathcal{L}_Xf(\varphi)=\lim_{h \to
  0}\frac{f(\varphi+h\,X(\varphi))-f(\varphi)}{h},\quad \varphi \in
\Diff.
\end{equation*}
If $U \subset \Diff$ is open and $X,Y:U \to C^\infty(\Circle)$ are
smooth, let
\begin{equation*}
D_XY(\varphi)=\lim_{h \to
  0}\frac{Y(\varphi+h\,X(\varphi))-Y(\varphi)}{h},\quad \varphi \in
\Diff.
\end{equation*}
This leads to a covariant definition of the Lie bracket of $X,Y
\in \mathcal{X}(\Diff),$
\begin{equation*}
\mathcal{L}_XY=[X,Y]=D_XY-D_YX.
\end{equation*}

Note that if $\mathcal{X}^R(\Diff)$ is the space of all
right-invariant smooth vector fields\footnote{~$X \in
\mathcal{X}^R(\Diff)$ is determined by its value $u$ at $Id$,
$X_{\eta}=R_{\eta\ast} u$ for $\eta \in \Diff$.} on $\Diff$, then
the bracket $[X,Y]$ of $X,\,Y \in \mathcal{X}^R(\Diff)$ is a
right-invariant vector field and $[X,Y]_{Id}=[u,v]$, where
$u=X_{Id},\, v=Y_{Id}$ cf. \cite{Mil}. This feature explains the
minus sign entering formula \eqref{LieBracket} - the commutation
operation is defined by this construction carried out with
right-invariant vector fields.


\section{Riemannian structures on $\Diff$}

We define an inner product on the Lie algebra $T_{Id}\Diff \equiv
C^\infty(\Circle)$ of $\Diff$, and extend it to $\Diff$ by
right-translation. The resulting right-invariant metric on $\Diff$
will be a weak Riemannian metric. In this section we discuss the
existence of the geodesic flow associated with this metric.

Consider on $T_{Id}\Diff \equiv C^{\infty}(\Circle)$ the
$H^k(\Circle)$ inner product\footnote{~ $H^k(\Circle)$, $k \ge 0$,
is the space of all $L^2(\Circle)$-functions (square integrable
functions) $f$ with distributional derivatives up to order $k$,
$\partial_x^i f$ with $i=0, \dotsc ,k$, in $L^2(\Circle)$. Endowed
with the norm
\begin{equation*}
\norm{ f }_k^2=\sum_{i=0}^{k}\int_{\Circle} (\partial_x^i
f)^2(x)\,dx,
\end{equation*}
$H^k(\Circle)$ becomes a Hilbert space. Note that if $\{
\hat{f}(j) \}_{j \in \mathbb{Z}}$ is the Fourier series of $ f \in
H^k(\Circle)$, then
\begin{equation*}
\norm{ f }_k^2=\sum_{j \in \mathbb{Z}} \Bigl( 1+ (2\pi j)^2 +
\dotsb +(2\pi j)^{2k}\Bigl) \abs{ \hat{f}(j) }^2.
\end{equation*}
}
\begin{equation*}
\langle u,\, v \rangle_k =\sum_{i=0}^{k}\int_{\Circle}
(\partial_x^i u)\,(\partial_x^i v)\,dx,\quad u,v \in
C^{\infty}(\Circle),
\end{equation*}
and extend this inner product to each tangent space $T_\eta\Diff,\
\eta \in \Diff$, by right-translation, i.e.
\begin{equation}\label{HkMetric}
\langle V,W \rangle_k:=\Big\langle
R_{\eta^{-1}\ast}V,\,R_{\eta^{-1}\ast} W \Big\rangle_k,\quad V,W
\in T_\eta{\Diff}.
\end{equation}
We have thus endowed $\Diff$ with a smooth right-invariant metric.
Note that the right-invariant metric \eqref{HkMetric} defines a
weak topology on $\Diff$ so that the existence of a covariant
derivative which preserves the inner product \eqref{HkMetric} is
not ensured on general grounds cf. \cite{Lan}. We will give a
constructive proof of the existence of such a covariant
derivative. Let us first note that
\begin{equation*}
\langle u,\, v \rangle_k =\int_{\Circle} A_k(u)\, v\, dx, \quad
u,v \in H^{k}(\Circle),\qquad k \ge 0,
\end{equation*}
where for every $n \ge 0$, $A_k: H^{n+2k}(\Circle) \to
H^{n}(\Circle)$ is the linear continuous isomorphism
\begin{equation}\label{InertiaOperator}
A_k=1-\frac{d^2}{dx^2}+ \dotsb +(-1)^k \frac{d^{2k}}{dx^{2k}}.
\end{equation}
This enables us to define the bilinear operator $B_k:
C^{\infty}(\Circle) \times C^{\infty}(\Circle) \to
C^{\infty}(\Circle)$,
\begin{equation}\label{BkOperator}
B_k(u,v)=-\,A_k^{-1}\Bigl( 2v_x A_k(u)+vA_k(u_x)\Bigr),\qquad u,v
\in  C^{\infty}(\Circle),
\end{equation}
with the property that
\begin{equation*}
\langle B_k(u,v),\, w \rangle_k=\langle u,\, [v,w]
\rangle_k,\qquad u,v,w \in  C^{\infty}(\Circle).
\end{equation*}
We can extend $B_k$ to a bilinear map $B_k$ on the space $
\mathcal{X}^R(\Diff)$ of smooth right-invariant vector fields on
$\Diff$ by
\begin{equation*}
B_k(X,Y)_\eta=R_{\eta\ast} B_k(X_{Id},Y_{Id}),\quad \eta \in
\Diff,\ X,Y \in \mathcal{X}^R(\Diff).
\end{equation*}
For $X \in \mathcal{X}(\Diff)$, let us denote by $X_{\eta}^R$ the
smooth right-invariant vector field on $\Diff$ whose value at
$\eta$ is $X_\eta$.

\begin{thm}\label{theorem1}
Let $k \ge 0$. There exists a unique Riemannian connection
$\nabla^k$ on $\Diff$ associated to the right-invariant metric
\eqref{HkMetric}, with
\begin{equation*}
(\nabla^k_XY)_\eta=[X,Y-Y_\eta^R]_\eta+\frac{1}{2}\,
\Bigl([X_\eta^R,Y_\eta^R]_\eta-B_k(X_\eta^R,Y_\eta^R)_\eta-
B_k(Y_\eta^R,X_\eta^R)_\eta\Bigr),
\end{equation*}
for smooth vector fields $X,Y$ on $\Diff$.
\end{thm}

\begin{proof}
The uniqueness of $\nabla^k$ is obtained like in
classical Riemannian geometry (see e.g. \cite{Lan}) and all the
required properties can be checked from its explicit
representation, using the defining identity for
$B_k$.
\end{proof}

The existence of $\nabla^k$ enables us to define parallel
translation along a curve on $\Diff$ and to derive the geodesic
equation of the metric defined by \eqref{HkMetric}. Throughout the
discussion, let $J \subset \mathbb{R}$ be an open interval with $0
\in J$. For a $C^1$-curve $\alpha:J \to \Diff$, let
$\hbox{Lift}(\alpha)$ be the set of lifts of $\alpha$ to $T\Diff$.
The derivation $D_{\alpha_t} :\hbox{Lift}(\alpha) \to
\hbox{Lift}(\alpha)$ along $\alpha$ is given in local coordinates
by
\begin{equation}\label{DerivationAlongCurve}
D_{\alpha_t}\gamma=\gamma_t-Q_k(\alpha_t\circ \alpha^{-1},\gamma
\circ \alpha^{-1})\circ \alpha,\qquad \gamma \in
\hbox{Lift}(\alpha),
\end{equation}
where $Q_k:C^\infty(\Circle) \times C^\infty(\Circle) \to
C^\infty(\Circle)$ is the bilinear operator
\begin{equation*}
Q_k(u,v)=\frac{1}{2}\,\Bigl( u_xv+uv_x+
B_k(u,v)+B_k(v,u)\Bigr),\quad u,v \in C^\infty(\Circle).
\end{equation*}
For a $C^1$-curve $\alpha:J \to \Diff$ we have
\begin{equation}\label{MetricDerivative}
\frac{d}{dt}\, \langle \gamma_1,\gamma_2\rangle_k=\langle
D_{\alpha_t}\gamma_1, \gamma_2\rangle_k + \langle
\gamma_1,D_{\alpha_t}\gamma_2\rangle_k,\qquad t \in J,
\end{equation}
for all $\gamma_1,\gamma_2 \in \hbox{Lift}(\alpha)$.

If $\alpha:J \to \Diff$ is a $C^2$-curve, a lift $\gamma:J \to
T\Diff$ is called $\alpha$-parallel if $D_{\alpha_t}\gamma \equiv
0$ on $J$. This is equivalent to requiring that
\begin{equation}\label{ParallelTransport}
v_t=\frac{1}{2}\, \Bigl( vu_x-v_xu+B_k(u,v)+B_k(v,u) \Bigr),
\end{equation}
where $u,v \in C^1(J;C^\infty (\Circle))$ are defined as $\alpha_t
\circ \alpha^{-1}=u$, respectively $\gamma \circ \alpha^{-1}=v$. A
$C^2$-curve $\varphi:J \to \Diff$ satisfying
$D_{\varphi_t}\varphi_t \equiv 0$ on $J$ is called a {\bf
geodesic}. If $u=\varphi_t \circ \varphi^{-1} \in T_{Id}\Diff
\equiv C^\infty(\Circle)$, then $u$ satisfies the equation
\begin{equation}\label{EulerEquation}
u_t=B_k(u,u), \qquad t \in J.
\end{equation}
Equation \eqref{EulerEquation} is the geodesic equation
transported by right-translation to the Lie algebra $T_{Id}\Diff$.
In a local chart the geodesic equation is
\begin{equation*}
\varphi_{tt}=P_k(\varphi,\varphi_t),
\end{equation*}
where $P_k$ is an operator that will be specified in the proof of
Theorem \ref{theorem2}. Assuming for the moment the local
existence of geodesics on $\Diff$ for the metric \eqref{HkMetric},
proved below, let us derive a conservation law for the geodesic
flow\footnote{~For finite-dimensional Lie groups the geodesic flow
of a one-sided invariant metric the angular momentum is preserved
\cite{ArnKhe}. This is a consequence of the invariance of the
metric by the action of the group on itself, in view of Noether's
theorem. The same reasoning can be carried over to the present
infinite-dimensional case.}. Observe that any $v \in
C^\infty(\Circle) \equiv T_{Id}\Diff$ defines a one-parameter
group of diffeomorphisms $h^s:\Diff \to \Diff,\,
h^s(\varphi)=\varphi \circ \exp_L(sv)$, where $\exp_L$ is the
Lie-group exponential map. The metric being by construction
invariant under the action of $h^s$, Noether's theorem ensures
that if $g:T\Diff \to \mathbb{R}$ stands for the right-invariant
metric, then
\begin{equation*}
\displaystyle\frac{\partial
g}{\partial \varphi_t}(\varphi,\varphi_t)\Bigl[
\frac{dh^s(\varphi)}{ds}\Bigl|_{s=0}\Bigr]
\end{equation*}
is preserved along the geodesic curve $t \mapsto \varphi(t)$ with
$\varphi(0)=Id$ and $\varphi_t(0)=u_0 \in T_{Id}\Diff$. We compute
\begin{align*}
\frac{dh^s(\varphi)}{ds}\Bigl|_{s=0}=\varphi_x \cdot v,\qquad
\frac{\partial g}{\partial v}(\varphi,v)\,[w]= 2\,\langle v \circ
\varphi^{-1},w \circ \varphi^{-1}\rangle_k,
    \intertext{obtaining that}
\langle \varphi_t \circ \varphi^{-1},\varphi_x \circ \varphi^{-1}
\cdot v \circ \varphi^{-1}\rangle_k=\langle u_0,v\rangle_k,\qquad
v \in C^\infty(\Circle).
    \intertext{Therefore}
\int_{\Circle} A_k(u)\cdot\varphi_x \circ \varphi^{-1}\cdot v
\circ \varphi^{-1}\,dx=\int_{\Circle} A_k(u_0)\cdot v\,dx,\qquad v
\in C^\infty(\Circle),
\end{align*}
where, as before, $u=\varphi_t \circ \varphi^{-1}$. A change of
variables yields
\begin{equation*}
\int_{\Circle} A_k(u)\circ\varphi \cdot \varphi_x^2 \cdot
v\,dx=\int_{\Circle} A_k(u_0)\cdot v\,dx,\qquad v \in
C^\infty(\Circle)
\end{equation*}
so that, denoting
\begin{equation}\label{Momentum}
m_k=A_k(u) \circ \varphi \cdot \varphi_x^2,
\end{equation}
we obtain
\begin{equation}\label{ConservationOfMomentum}
m_k(t)=m_k(0),\qquad t \in [0,T),
\end{equation}
where $\varphi \in C^2([0,T);\Diff)$ is the geodesic for the
metric \eqref{HkMetric}, starting at $\varphi(0)=Id \in \Diff$ in
the direction $u_0=\varphi_t(0) \in T_{Id}\Diff$; as before,
$u=\varphi_t \circ \varphi^{-1}$.

To prove the existence of geodesics, we proceed as follows. The
classical local existence theorem for differential equations with
smooth right-hand side, valid for Hilbert spaces (see \cite{Lan}),
does not hold in $C^\infty (\Circle)$ cf. \cite{Ham}. However,
note that $C^\infty (\Circle)= \displaystyle\cap_{n \ge 2k+1} H^n(
\Circle)$. We use the classical approach to prove that for every
$n \ge 2k+1$, the geodesic equation has, on some maximal interval
$[0,T_n)$ with $T_n>0$, a unique solution in $H^n( \Circle)$,
depending smoothly on time. A priori $T_n \le T_{2k+1}$. It turns
out that $T_n = T_{2k+1}$ for all $n \ge 2k+1$, fact that will
ensure the existence of geodesics on $\Diff$ endowed with the
right-invariant metric \eqref{HkMetric} for every $k \ge 1$. The
peculiarities of the special case $k=0$ (where this approach is
not applicable) are discussed in the last section.\bigskip

\begin{thm}\label{theorem2}
Let $k \ge 1$. For every $u_0 \in C^\infty (\Circle)$, there
exists a unique geodesic $\varphi \in C^\infty([0,T);\Diff)$ for
the metric \eqref{HkMetric}, starting at $\varphi(0)=Id \in \Diff$
in the direction $u_0=\varphi_t(0) \in T_{Id}\Diff$. Moreover, the
solution depends smoothly on the initial data $u_0\in C^\infty
(\Circle)$.
\end{thm}

\begin{proof}
Note that
\begin{equation*}
uA_k(u_x)=A_k(uu_x)+C_k^0(u),\quad u \in H^n(\Circle),\ n \ge
2k+1,
\end{equation*}
where $C_k^0:H^n(\Circle) \to H^{n-2k}(\Circle)$ is a
$C^\infty$-operator depending quadratically on $u,u_x,\dotsc
,\partial_x^{2k}u$. Denoting by $C_k:H^n(\Circle) \to
H^{n-2k}(\Circle)$ the $C^\infty$-operator
\begin{equation*}
C_k(u)=-C_k^0(u)-2u_xA_k(u),
\end{equation*}
we obtain that
\begin{equation*}
B_k(u,u)=A_k^{-1}C_k(u)-uu_x,\quad u \in H^n(\Circle),\ n \ge
2k+1.
\end{equation*}
The geodesic equation \eqref{EulerEquation} becomes
\begin{equation*}
u_t+uu_x=A_k^{-1}C_k(u),
\end{equation*}
where $u=\varphi_t \circ \varphi^{-1} \in C^\infty(\Circle)$.
Letting $v=u \circ \varphi=\varphi_t$, we can write the geodesic
equation in a local chart $U$ in $C^\infty(\Circle)$ as
\begin{equation}\label{Cauchy}
  \left\{ \begin{aligned}
     \varphi_t & = v, \\
     v_t & = P_k(\varphi,v),
  \end{aligned}
\right.
\end{equation}
where
\begin{equation*}
P_k(\varphi,v)=\Bigl[A_k^{-1}C_k(v \circ \varphi^{-1})\Bigr]\circ
\varphi.
\end{equation*}
The operator
\begin{equation*}
(\varphi,v) \mapsto (\varphi,\,P_k(\varphi,v))
\end{equation*}
can be decomposed into $Q_k\circ E_k$ with
\begin{align*}
E_k(\varphi,v)=\Bigl(\varphi,\, R_{\varphi} \circ C_k \circ
R_{\varphi^{-1}}(v) \Bigr),
\intertext{and}
Q_k(\varphi,v)=\Bigl(\varphi,\,R_{\varphi} \circ A_k^{-1} \circ
R_{\varphi^{-1}}(v) \Bigr).
\end{align*}
Specifying the explicit form of $E_k(\varphi,v)$, we see that this
operator extends to the space $U_n \times H^n( \Circle)$, where
$U_n$ is the open subset of $H^n( \Circle)$ of all functions
having a strictly positive derivative\footnote{~Note that $n \ge
2k+1 \ge 3$ so that $H^n( \Circle)$-functions are of class $C^2$.
Explicit calculations show that if $\eta \in U_n$, then $\eta$ is
a $H^n( \Circle)$-homeomorphism of the circle with $\eta^{-1} \in
H^n( \Circle)$.}. The same argument can be pursued in the case of
the operator $G_k:U_n \times H^n( \Circle) \to U_n \times
H^{n-2k}( \Circle)$,
\begin{equation*}
G_k(\varphi,v)=\Bigl(\varphi,\,R_{\varphi} \circ A_k \circ
R_{\varphi^{-1}}(v) \Bigr),
\end{equation*}
the inverse of $Q_k$ (as a map). Direct calculations confirm that
$E_k$ and $G_k$ are both smooth maps from $U_n \times H^n(
\Circle)$ to $U_n \times H^{n-2k}( \Circle)$. The regularity of
$G_k$ ensures that its Fr{\'e}chet differential can be computed by
calculating directional derivatives. One finds that
\begin{equation*}
DG_k{(\varphi,v)}=\left( \begin{array}{cc}
    Id & 0 \\
    \ast & \sum_{i=0}^{2k} a_i(\varphi)\,\partial_x^i
\end{array} \right) ,
\end{equation*}
with
\begin{equation*}
 a_{2k}=\frac{(-1)^k}{\varphi_x^{2k}}
\end{equation*}
and $a_0,\dotsc ,a_{2k-1},$ all of the form
\begin{equation*}
 \frac{p_k(\varphi,\varphi_x,\dotsc ,\partial_x^{2k}\varphi)}{\varphi_x^{4k}}
\end{equation*}
for a polynomial $p_k$ with constant coefficients, while $\ast$ is
a linear differential operator of order $2k$ with coefficients
rational functions of the form
\begin{equation*}
\frac{q(\varphi,v,\varphi_x,v_x,\dotsc ,\partial_x^{2k}\varphi,
\partial_x^{2k}v)}{\varphi_x^{4k}}
\end{equation*}
for some constant coefficient polynomial $q$.
Observe\footnote{~This can be proved using the Fourier
representation of functions in $H^j(\Circle),\, j \ge 0$.} that
for every $f \in H^{n-2k}(\Circle)$ there is a unique solution $u
\in H^{n}(\Circle)$ of the ordinary linear differential equation
with $H^{n-2k}(\Circle)$-coefficients
\begin{equation*}
\sum_{i=0}^{2k} a_i(\varphi)\,\partial_x^i v=f.
\end{equation*}
Taking into account the form of $DG_k$, we infer that
\begin{equation*}
DG_k{(\varphi,v)} \in \hbox{Isom}\Bigl(U^n \times
H^n(\Circle),\,U^n \times H^{n-2k}(\Circle)\Bigr).
\end{equation*}
Since the differential of the smooth map $G_k: U_n \times
H^n(\Circle) \to U_n \times H^{n-2k}( \Circle)$ is invertible at
every point, from the inverse function theorem on Hilbert spaces
\cite{Lan} we deduce that its inverse $Q_k:U_n \times
H^{n-2k}(\Circle) \to U_n \times H^n(\Circle)$ is also smooth. The
regularity properties that we just proved for the maps $Q_k$,
$E_k$, show that $P_k$ is a smooth map from $U_n\times
H^n(\Circle)$ to $H^n(\Circle)$.

Regard \eqref{Cauchy} as an ordinary differential equation on $U_n
\times H^n(\Circle)$, with a smooth right-hand side, viewed as a
map from $U_n \times H^n(\Circle)$ to $U_n \times H^n(\Circle)$.
The Lipschitz theorem for differential equations in Banach spaces
\cite{Lan} ensures that for every ball $B(0,\varepsilon_n)\subset
H^n(\Circle)$ there exists $T_n=T_n(\varepsilon_n)>0$ such that
for every $u_0 \in B(0,\varepsilon_n)$, the equation
\eqref{Cauchy} with data $\varphi(0) = Id$ and $v(0) = u_0$ has a
unique solution $(\varphi,v)\in C^{\infty}([0, T_n);U_n\times
H^n(\Circle))$. Moreover, this solution $(\varphi,v)$ depends
smoothly on the initial data $u_0$ and can be extended to some
maximal existence time $T_n^\ast>0$. If $T_n^\ast<\infty$, we have
either that $\limsup_{t \uparrow T_n^\ast}\,\norm{v(t)}_n=\infty$
or there is a sequence $t_j \uparrow T_n^\ast$ such that
$\varphi(t_j)$ accumulates at the boundary of $U_n$ as $j \to
\infty$.

Choose some ball $B(0,\varepsilon_{2k+1})\subset
H^{2k+1}(\Circle)$. We prove now that for any $ u_0 \in
B(0,\varepsilon_{2k+1}) \,\cap \, C^\infty(\Circle)$ there exists
a unique geodesic $\varphi \in C^\infty([0,T_{2k+1});\Diff)$ for
the metric \eqref{HkMetric}, starting at $\varphi(0)=Id$ in the
direction $u_0$. Since $u_0 \in H^n( \Circle)$ for every $n \ge
2k+1$, it suffices to prove that the solution $(\varphi,v)$ of
equation \eqref{Cauchy} on each $U_n \times H^n(\Circle)$, with
data $\varphi(0) = Id$ and $v(0) = u_0$, has the maximal existence
time $T_n=T_{2k+1}$. Assuming that $T_{2k+2}<T_{2k+1}$, note that
$(\varphi(T_{2k+2}),v(T_{2k+2}))$ is defined in $U_{2k+1}\times
H^{2k+1}(\Circle)$ and $\varphi(T_{2k+2})$ is a
$C^1$-diffeomorphism of the circle. Recall the notation $u = v
\circ \varphi^{-1}$.

To prove that $\varphi(t)$ converges in $U_{2k+2}(\Circle)$ as $t
\uparrow T_{2k+2}$, let us use $\varphi_t=u \circ \varphi$ to
compute $\partial_x^{2k}\varphi_t$, $t \in (0,T_{2k+1})$. We
obtain
\begin{equation*}
\varphi_x \cdot \partial_x^{2k}\varphi_t - \varphi_{tx} \cdot
\partial_x^{2k}\varphi= (-1)^k\varphi_x^2\,\Bigl[
\varphi_x^{2k-3}m_k(t)+ \mathcal{E}_k(v,\varphi)\Bigr],\quad t \in
(0,T_{2k+1}),
\end{equation*}
where $\mathcal{E}_k(v,\varphi)$ is a smooth expression containing
only $x$-derivatives of $\varphi$ of order $i \le 2k-1$ and
$x$-derivatives of $v$ of order $j \le 2k-1$. Hence
\begin{align*}
    \frac{d}{dt}\Bigl(\frac{\partial_x^{2k}\varphi}{\varphi_x}\Bigr)
        & = (-1)^k\,\Bigl[\varphi_x^{2k - 3}\cdot m_k(t) + \mathcal{E}_k(v,\varphi)\Bigr] \\
    & = (-1)^k\,\Bigl[\varphi_x^{2k - 3}\cdot m_k(0) + \mathcal{E}_k(v,\varphi)\Bigr],
        &\qquad t \in (0,T_{2k+1}),
\end{align*}
in view\footnote{~Relation \eqref{ConservationOfMomentum} was
derived assuming the existence of geodesics on $\Diff$. In the
present context, it can be proved as follows. Define $m_k$ by
\eqref{Momentum} and note that it is a polynomial expression in
$v,\varphi,v_x,\varphi_x,\dotsc
,\partial_x^{2k}v,\partial_x^{2k}\varphi,$ divided by some power
of $\varphi_x$. Therefore $m_k \in
C^\infty([0,T_{2k+1});H^1(\Circle))$ and, using \eqref{Cauchy}, it
can be checked by differentiation with respect to $t$ that
$m_k(t)=m_k(0)$ for all $t \in [0,T_{2k+1})$.} of
\eqref{ConservationOfMomentum}. For $t \in (0,T_{2k+1})$ we obtain
that
\begin{equation}\label{PhiDerivative}
\partial_x^{2k}\varphi(t)=(-1)^k\,\varphi_x(t)\,\int_0^t
\Bigl[\varphi_x^{2k - 3}\cdot m_k(0) +\mathcal{E}_k(v,\varphi)\Bigr]\,ds.
\end{equation}
Since $m_k(0) \in C^\infty(\Circle)$ and
\begin{multline*}
    (\varphi,v) \in C^\infty([0,T_{2k+1});U_{2k+1} \times
H^{2k+1}(\Circle))\\
    \cap \ C^\infty([0,T_{2k+2});U_{2k+2} \times
H^{2k+2}(\Circle)),
\end{multline*}
differentiating \eqref{PhiDerivative} twice with respect to $x$,
we infer that $(\varphi(t),\varphi_t(t))$ converges in
$U_{2k+2}(\Circle) \times H^{2k+2}(\Circle)$ as $t \uparrow
T_{2k+2}$. The limit can only be
$(\varphi(T_{2k+2}),v(T_{2k+2}))$. Therefore $T_{2k+2}=T_{2k+1}$.
This procedure can be repeated for $n=2k+3$ etc. and the existence
of the smooth geodesics on $\Diff$ is now plain.
\end{proof}

The previous results enable us to define the Riemannian
exponential map $\mathfrak{exp}$ for the $H^k$ right-invariant
metric ($k\geq 1$). If $\varphi(t;u_0)$ is the geodesic on
$\Diff$, starting at $Id$ in the direction $u_0 \in C^\infty(
\Circle)$, note the homogeneity property
\begin{equation}\label{HomogeneityProperty}
\varphi(t;su_0)=\varphi(ts;u_0)
\end{equation}
valid for all $t,\, s \ge 0$ such that both sides are
well-defined. In the proof of Theorem \ref{theorem2} we saw that
there exists $\delta>0$ so that all geodesics $\varphi(t;u_0)$ are
defined on the same time interval $[0,T]$ with $T>0$, for all $u_0
\in \Diff$ with $\norm{ u_0 }_{2k+1}<\delta$. Hence, we can define
$\mathfrak{exp}(u_0)=\varphi(1;u_0)$ on the open set
\begin{equation*}
\set{u_0 \in  \Diff;\quad \norm{ u_0 }_{2k+1} <
\frac{2\,\delta}{T} }
\end{equation*}
of $\Diff$, and the map $u_0 \mapsto \mathfrak{exp}(u_0)$ is
smooth.

\begin{thm}\label{theorem3}
The Riemannian exponential map for the $H^k$ right-invariant
metric on $\Diff$, $k \ge 1$, is a smooth local diffeomorphism
from a neighborhood of zero on $T_{Id}\Diff$ to a neighborhood of
$Id$ on $\Diff$.
\end{thm}

Let us first establish

\begin{lem}\label{lemma1}
Let $n \ge 2k+1$ and let $(\varphi,v)$ be a solution of
\eqref{Cauchy} with data $(Id,u_0)\in U_n \times H^n( \Circle)$,
defined on $[0,T)$. If there exists $t \in [0,T)$ such that
$\varphi (t) \in U_{n+1}$ then $u_0 \in H^{n+1}( \Circle)$.
\end{lem}

\begin{proof}
From \eqref{PhiDerivative} we get
\begin{equation*}
\frac{\partial_x^{2k}
\varphi(t)}{\varphi_x(t)}=(-1)^k\,\,m_k(0)\cdot\int_0^t
\varphi_x^{2k - 3}\,ds  + (-1)^k\,\int_0^t
\mathcal{E}_k(v,\varphi)\,ds,\quad t \in [0,T).
\end{equation*}
If for some $t \in [0,T)$ we have $\varphi (t) \in H^{n+1}(
\Circle)$, the fact that $\varphi_x$ is strictly positive forces
$m_k(0)\in H^{n+1-2k}( \Circle)$ and therefore $u_0\in H^{n+1}(
\Circle)$.
\end{proof}

\begin{proof}[Proof of Theorem \ref{theorem3}.]
Viewing $\mathfrak{exp}$ as a smooth map from a small neighborhood
of $0 \in  H^n( \Circle)$ to $U_n$, $n \ge 2k+1$, its differential
at $0 \in  H^n( \Circle)$, $D\mathfrak{exp}_0$, is the identity
map. Indeed, for $v \in  H^n( \Circle)$ we have by
\eqref{HomogeneityProperty} that $\mathfrak{exp}(tv)=\varphi(t;v)$
so that
\begin{equation*}
\frac{d}{dt}\,\mathfrak{exp}(tv)\Bigl|_{t=0}=
\frac{d}{dt}\,\varphi(t;v)\Bigl|_{t=0}=v.
\end{equation*}
As a consequence of the inverse function theorem on Hilbert
spaces, we can find open neighborhoods $V_{2k+1}$ and $O_{2k+1}$
of $0 \in H^{2k+1}(\Circle)$ and $Id \in U_{2k+1}$, respectively,
such that $\mathfrak{exp}: V_{2k+1}\rightarrow O_{2k+1}$ is a
$C^{\infty}$ diffeomorphism with $D\mathfrak{exp}_{u_0}:
H^{2k+1}(\Circle) \to H^{2k+1}(\Circle)$ bijective for every $u_0
\in V_{2k+1}$. We already know from the proof of Theorem
\ref{theorem2} that
\begin{equation*}
\mathfrak{exp} \Bigl( V_{2k+1} \cap C^{\infty}( \Circle) \Bigr)
\subset O_{2k+1} \cap C^{\infty}( \Circle),
\end{equation*}
while Lemma \ref{lemma1} ensures that $\mathfrak{exp}$ is a local
bijection between these open sets. It remains to show that
$\mathfrak{exp}$ is a smooth diffeomorphism from $V_{2k+1} \cap
C^{\infty}( \Circle)$ to $O_{2k+1} \cap C^{\infty}( \Circle)$.

Let $u_0 \in V_{2k+1} \cap C^{\infty}( \Circle)$. We know that
$D\mathfrak{exp}_{u_0}$ is a bounded linear operator from $H^n(
\Circle)$ to $H^n( \Circle)$ for every $n \ge 2k+1$ and we will
prove that it is actually a bijection. Then, in view of the
inverse function theorem on Hilbert spaces, both $\mathfrak{exp}$
and its inverse are smooth maps on small
$H^n(\Circle)$-neighborhoods of $u_0 \in V_{2k+1} \cap C^{\infty}(
\Circle)$, respectively $\mathfrak{exp}(u_0) \in O_{2k+1} \cap
C^{\infty}( \Circle)$. Letting $n \uparrow \infty$, this would
show that $\mathfrak{exp}$ is locally a smooth diffeomorphism.

To prove this last step, we use an inductive argument. To start
with, $D\mathfrak{exp}_{u_0}$ is a bijection from
$H^{2k+1}(\Circle)$ to $H^{2k+1}(\Circle)$ as $u_0 \in V_{2k+1}$.
For a fixed $n \ge 2k+1$, assume that the map
$D\mathfrak{exp}_{u_0}$ is a bijection from $H^j( \Circle)$ to
$H^j( \Circle)$ for all $j=2k+1,\dotsc ,n,$ and let us show that
$D\mathfrak{exp}_{u_0}$ is a bijection from $H^{n+1}( \Circle)$ to
$H^{n+1}( \Circle)$. First of all, $D\mathfrak{exp}_{u_0}$ is
injective as a bounded linear map from $H^{n+1}( \Circle)$ to
$H^{n+1}( \Circle)$ since its extension to $H^{n}( \Circle)$ is
injective. To prove its surjectivity as a map from $H^{n+1}(
\Circle)$ to $H^{n+1}( \Circle)$, it suffices\footnote{~Since
$\mathfrak{exp}$ is a smooth map on $V_{2k+1} \cap H^{n+1}(
\Circle)$ we have $D\mathfrak{exp}_{u_0}(H^{n+1}( \Circle))
\subset H^{n+1}( \Circle)$ while the inductive assumption ensures
$D\mathfrak{exp}_{u_0}(H^{n}( \Circle))= H^{n}( \Circle)$.} to see
that there is no $w \in H^{n}( \Circle),\, w \not\in H^{n+1}(
\Circle),$ with $D\mathfrak{exp}_{u_0}(w) \in H^{n+1}( \Circle)$.
Assume there is such a $w$. For $\varepsilon>0$ small enough let
$\varphi^\varepsilon(t)$ be the solution of \eqref{Cauchy} on
$U_{n}$ starting at $Id$ in the direction $u_0+\varepsilon w$,
with the corresponding $v^\varepsilon \in H^{n}( \Circle))$. We
know that the map $(\varphi^\varepsilon(t), v^\varepsilon(t)) \in
U_n \times H^{n}( \Circle)$ depends smoothly on $\varepsilon$ and
$t \in [0,1]$. From \eqref{PhiDerivative} we obtain
\begin{multline*}
    (-1)^k\,\frac{\partial_x^{2k}
\varphi^\varepsilon(1)}{\varphi_x^\varepsilon(1)} =
\Bigl(m_k(0;u_0) +\varepsilon m_k(0;w)\Bigr)\,\int_0^1
(\varphi_x^\varepsilon)^{2k - 3}\,ds \\
   +\, \int_0^1
\mathcal{E}_k(v^\varepsilon,\varphi^\varepsilon)\,ds.
\end{multline*}
Differentiating with respect to $\varepsilon$, a calculation shows
that
\begin{equation*}
D\mathfrak{exp}_{u_0}(w)=\dfrac{d}{d\varepsilon}\,
\varphi^\varepsilon (1)\Bigl|_{\varepsilon=0} \in H^{n+1}(
\Circle)
\end{equation*}
is possible only if $m_k(0;w) \in H^{n-2k+1}( \Circle)$, i.e. $w
\in H^{n+1}( \Circle)$. The obtained contradiction concludes the
proof.
\end{proof}


\section{Minimizing property of the geodesics}

Throughout this section we prove the length minimizing property
for the geodesics of the right-invariant metric \eqref{HkMetric}
on $\Diff$ for some fixed $k \ge 1$.

Let $V_0$ be a vector tangent at $\alpha(0)=\alpha_0$ to a
$C^2$-curve $\alpha:J \to \Diff$. The parallel transport of $V$
along the curve $\alpha$ is defined as a curve $\gamma \in
\hbox{Lift}(\alpha)$ with $\gamma(0)=V_0$ and $D_{\alpha_t}\gamma
\equiv 0$ on $J$.

\begin{lem}\label{lemma2}
Let $\alpha:J \to \Diff$ be a $C^2$ curve. Given $V_0 \in
T_{\alpha_0}\Diff,\, \alpha_0=\alpha(0) \in \Diff$, there exists a
unique lift $\gamma: J \to T\Diff$ which is $\alpha$-parallel and
such that $\gamma(0)=V_0$. Moreover, if $\gamma_1,\, \gamma_2$ are
the unique $\alpha$-parallel lifts of $\alpha$ with
$\gamma_i(0)=V_i \in T_{\alpha_0}\Diff,\, i=1,2$, then
\begin{equation*}
\langle \gamma_1(t),\,\gamma_2(t)\rangle_k=\langle V_1,V_2\rangle_k,\qquad t
\in J.
\end{equation*}
\end{lem}

\begin{proof}
In view of \eqref{BkOperator} and \eqref{ParallelTransport}, the
equation of parallel transport is
\begin{equation*}
v_t=\frac{1}{2}\,(vu_x-uv_x)\, - \,A_k^{-1}\Bigl[v_xA_k(u)+u_xA_k(v)+
\frac{1}{2}\,vA_k(u_x)+ \frac{1}{2}\,uA_k(v_x)\Bigr],
\end{equation*}
where $u=\alpha_t \circ \alpha^{-1}$ and $v=\gamma \circ
\alpha^{-1}$. Note that the operators
\begin{equation*}
      (u,v)  \mapsto A_k^{-1}[v_xA_k(u)+u_xA_k(v)],
\end{equation*}
and
\begin{equation*}
(u,v)  \mapsto   \frac{1}{2}\,
A_k^{-1}\Bigl[vA_k(u_x)+uA_k(v_x)\Bigr]\,-\frac{1}{2}\,(vu_x+uv_x)
\end{equation*}
are smooth from $H^n( \Circle) \times H^n( \Circle)$ to $H^n(
\Circle)$ for every $n \ge 2k+1$. Denote by $\Theta_k(u,v)$ their
sum. The equation of parallel transport can be written as
\begin{equation}\label{EquationOfParallelTransport}
v_t+uv_x+\Theta_k(u,v)=0.
\end{equation}
For a fixed $u \in C^1(J;\Diff)$, the map $v \mapsto
\Theta_k(u,v)$ is a bounded linear operator from $H^n( \Circle)$
to $H^n( \Circle)$ for every $n \ge 2k+1$. Viewing
\eqref{EquationOfParallelTransport} as linear hyperbolic evolution
equation in $v$ with fixed $u \in C^1(J;\Diff)$, it is known (see
\cite{Kat}) that, given $V_0 \in H^n( \Circle),\,n \ge 2k+1$,
there exists a unique solution
\begin{equation*}
v \in C(J;\,H^{n}( \Circle)) \cap \, C^1(J;H^{n-1}(\Circle))
\end{equation*}
of \eqref{EquationOfParallelTransport} with initial data
$v(0)=V_0$. Letting $n \uparrow \infty$, we infer that, given $V_0
\circ \alpha_0^{-1} \in T_{Id}\Diff \equiv C^\infty(\Circle)$,
there exists a unique solution $v \in C^1(J;\Diff)$ to
\eqref{EquationOfParallelTransport} with $v(0)= V_0 \circ
\alpha_0^{-1}$.

From \eqref{MetricDerivative} we deduce that $\langle
\gamma_1(t),\,\gamma_2(t)\rangle_k$ is constant for any
$\alpha$-parallel lifts and the second assertion follows.
\end{proof}

Choose open neighborhoods $\mathcal{W}$ of $0 \in
C^\infty(\Circle)$, respectively $\mathcal{U}$ of $Id \in \Diff$,
such that $D\mathfrak{exp}_{u_0}: H^{2k+1}(\Circle) \to
H^{2k+1}(\Circle)$ is bijective for every $u_0 \in \mathcal{W}$
and $\mathfrak{exp}$ is a smooth diffeomorphism from $\mathcal{W}$
onto $\mathcal{U}$, cf. Theorem \ref{theorem3}. The map
\begin{equation*}
G:\Diff \times \mathcal{W} \to \Diff \times \Diff, \quad (\eta,u)
\mapsto \Bigl(\eta, R_\eta\, \mathfrak{exp}(u)\Bigr),
\end{equation*}
is a smooth diffeomorphism onto its image. Let
$\mathcal{U}(\eta)=R_\eta\, \mathcal{U}=R_\eta\, \mathfrak
{exp}(\mathcal{W})$. If $\varphi \in \mathcal{U}(\eta)-\{\eta\}$,
then $\varphi=\mathfrak{exp}(v) \circ \eta$ for some $v \in
\mathcal{W}$. Let $v=r\, w$, where $\langle w,w\rangle_k=1$ and $r
\in \mathbb{R}_+$ to define the polar coordinates $(r,w)$ of
$\varphi \in \mathcal{U}(\eta)$.

If $\sigma:J_1 \times J_2 \to \Diff$ is a map such that
$\dfrac{\partial^2 \sigma}{\partial r^2}$, $\dfrac{\partial^2
\sigma}{\partial t\,\partial r}$ and $\dfrac{\partial^2
\sigma}{\partial r\,\partial t}$ are continuous, denote by
$\partial_1 \sigma$ the partial derivative with respect to $r$ and
define similarly $\partial_2 \sigma$. Both curves $r \mapsto
\partial_1\sigma(r,t)$ and $r \mapsto \partial_2\sigma(r,t)$ are
lifts of $r \mapsto \sigma(r,t)$. Generally, if $\gamma$ is a lift
of $r \mapsto \sigma(r,t)$, let
$(D_1\gamma)(r,t)=(D_{\partial_1\sigma}\gamma)(r)$ and define
$D_2\gamma$ similarly. In a local chart we have by
\eqref{DerivationAlongCurve} that
\begin{equation}\label{SchwarzRule}
D_1\partial_2\sigma=\partial_1\partial_2\sigma-Q_k(\partial_1\sigma,
\partial_2\sigma)=D_2\partial_1\sigma
\end{equation}
since $Q_k$ is symmetric. On the other hand, from
\eqref{MetricDerivative} we infer
\begin{equation*}
\partial_2\, \langle \, \partial_1\sigma,\, \partial_1\sigma\,\rangle_k=
2\,  \langle\, D_2\partial_1\sigma,\,\partial_1\sigma\,\rangle_k.
\end{equation*}
The previous relation combined with \eqref{SchwarzRule} yields
\begin{equation}\label{SecondDerivative}
\partial_2\, \langle \, \partial_1\sigma,\, \partial_1\sigma\,\rangle_k=
2\, \langle\, D_1\partial_2\sigma,\,\partial_1\sigma\,\rangle_k.
\end{equation}

\begin{lem}\label{lemma3}
Let $\gamma:[a,b] \to \mathcal{U}(\eta)-\{\eta\}$ be a piecewise $C^1$-curve. Then
\begin{equation*}
l(\gamma) \ge \abs{ r(b)-r(a) },
\end{equation*}
where $l(\gamma)$ is the length of the curve and $(r(t),w(t))$ are
the polar coordinates of $\gamma(t)$. Equality holds if and only
if the function $t \mapsto r(t)$ is monotone and the map $t
\mapsto w(t) \in \mathcal{W}$ is constant.
\end{lem}

\begin{proof}
We may assume without loss of generality that $\gamma$ is $C^1$
(in the general case, break $\gamma$ up into pieces that are
$C^1$) and that $\eta=Id$ (in view of the right-invariance
property of the metric). Observe that $w(t)$ is obtained in a
chart by the inversion of $\mathfrak{exp}$ followed by a
projection so that the functions $t \mapsto r(t)$ and $t \mapsto
w(t)$ are of class $C^1$.

Let $\sigma(r,t)=\mathfrak{exp}(r\, w(t))$. Let $\varphi(s;z)$ be
the solution of \eqref{Cauchy} starting at $Id$ in the direction
$z \in C^\infty(\Circle)$. Relation \eqref{HomogeneityProperty}
yields $\sigma(r,t)=\varphi(r;w(t))$, while the proof of Theorem
\ref{theorem2} ensures for every $n \ge 2k+1$ the smooth
dependence of $\varphi(s;z)$ on $s$ as well as the smooth
dependence of $(\varphi,\varphi_s)$ on $z$ in $H^n(\Circle)$.
Therefore $\dfrac{
\partial^2\sigma}{\partial r^2}$ and $\dfrac{\partial^2\sigma}
{\partial t\,\partial r}$ are continuous in the
$H^n(\Circle)$-setting for every $n \ge 2k+1$. Furthermore, since
\begin{equation*}
\varphi(s;z)=Id+\int_0^s \frac{\partial\varphi}{\partial s}
(\xi;z)\, d\xi\quad\hbox{in} \quad H^n(\Circle),
\end{equation*}
we have
\begin{equation*}
\frac{\partial\varphi}{\partial z}(s;z)=\int_0^s \frac{\partial^2\varphi}{
\partial z\partial s}(\xi;z)\, d\xi \quad\hbox{in}
\quad \mathcal{L}(H^n(\Circle),H^n(\Circle)),
\end{equation*}
thus $\displaystyle \frac{\partial^2\varphi}{
\partial z\partial s}=\frac{\partial^2\varphi}{
\partial s\partial z}$. But $t \mapsto w(t) \in H^n(\Circle)$
is a $C^1$-map so that $\dfrac{\partial^2\sigma}{\partial
r\,\partial  t}$ is also continuous in the $H^n(\Circle)$-setting
for every $n \ge 2k+1$. Letting $n \uparrow \infty$ we obtain that
$\dfrac{\partial^2\sigma} {\partial r^2},\,
\dfrac{\partial^2\sigma}{\partial r\,\partial t}$ and
$\dfrac{\partial^2\sigma}{\partial t\,\partial r}$ are all
continuous in the $C^\infty(\Circle)$-topology.

Note that
\begin{equation}\label{GammaDerivative}
\gamma'(t)=\frac{\partial\, \sigma}{\partial\, r}\cdot r'(t)\,+\,
\frac{\partial\, \sigma}{\partial\,t},\qquad t \in J.
\end{equation}
Since $r \mapsto \sigma(r,t)$ is a geodesic, we
obtain by Lemma \ref{lemma2} that
\begin{equation}\label{SigmaDerivative}
\langle\, \frac{\partial\, \sigma}{\partial\, r}\,,\,
\frac{\partial\, \sigma}{\partial\,
r}\,\rangle_k=\langle w(t),w(t)\rangle_k \equiv 1.
\end{equation}
Let us now show that
\begin{equation}\label{OrthogonalDerivatives}
\langle\, \frac{\partial\, \sigma}{\partial\, r}\,,\,
\frac{\partial\, \sigma}{\partial\,
t}\,\rangle_k \equiv 0.
\end{equation}
Indeed, from \eqref{SecondDerivative} and \eqref{SigmaDerivative}
we obtain that
\begin{equation*}
\langle\, D_1\,\frac{\partial\, \sigma}
{\partial\,
t}\,,\, \frac{\partial\, \sigma}{\partial\, r}\rangle_k=\frac{1}{2}\,
\partial_t\, \langle\,
 \frac{\partial\, \sigma}{\partial\, r}\,,\frac{\partial\, \sigma}{\partial\,
r}\,\rangle_k \equiv 0.
\end{equation*}
This, in combination with \eqref{MetricDerivative}, leads to
\begin{equation*}
\partial_r\, \langle\, \frac{\partial\, \sigma}{\partial\, r}\,,
\frac{\partial\, \sigma} {\partial\, t}\,\rangle_k = \langle\,
D_1\, \frac{\partial\, \sigma}{\partial\, r}\,, \,\frac{\partial\,
\sigma} {\partial\, t}\,\rangle_k+\langle\,  \frac{\partial\,
\sigma}{\partial\, r}\,,D_1\, \frac{\partial\, \sigma} {\partial\,
t}\,\rangle_k \equiv 0,
\end{equation*}
since $(D_1\, \frac{\partial\, \sigma}{\partial\, r})=0$ as $r
\mapsto \sigma(r,t)$ is a geodesic. The previous relation yields
\begin{equation*}
\langle \frac{\partial\, \sigma}{\partial\, r} \,,
\frac{\partial\, \sigma}{\partial\, t} \rangle_k(r,t)=\langle\,
\frac{\partial\, \sigma}{\partial\, r}\,, \frac{\partial\,
\sigma}{\partial\, t}\,\rangle_k(0,t).
\end{equation*}
But $\sigma(0,t)=Id$ forces $\dfrac{\partial\, \sigma}{\partial\,
r}(0,t)=0$ and therefore \eqref{OrthogonalDerivatives} holds.

Combining \eqref{GammaDerivative}-\eqref{OrthogonalDerivatives},
we obtain
\begin{equation*}
\norm{ \gamma'(t) }_k^2 = \abs{ r'(t) }^2 + \norm{
\frac{\partial\, \sigma}{\partial\, t} }_k^2 \ge \abs{ r'(t)
}^2,\qquad t \in [a,b],
\end{equation*}
so that the length of $\gamma$ is estimated by
\begin{equation*}
l(\gamma) \ge \int_a^b \abs{ r'(t) }\,dt\ge \abs{ r(b)-r(a) }.
\end{equation*}
Since $\norm{ \dfrac{\partial\, \sigma}{\partial\, t} }_k \equiv
0$ forces $w'(t)=0$ as $D\mathfrak{exp}_{rw(t)}$ is a bijection
from $H^{2k+1}(\Circle)$ to $H^{2k+1}(\Circle)$, the
characterization of the equality case follows at once.
\end{proof}

Let us now prove

\begin{thm}\label{theorem4}
If $\eta,\varphi \in \Diff$ are close enough, more precisely, if
$\varphi \circ \eta^{-1} \in \mathcal{U}$, then $\eta$ and
$\varphi$ can be joined by a unique geodesic in
$\mathcal{U}(\eta)$. Among all piecewise $C^1$-curves joining
$\eta$ to $\varphi$ on $\Diff$, the geodesic is length minimizing.
\end{thm}

\begin{proof}
Observe that if $v=\mathfrak{exp}^{-1}
(\varphi \circ \eta^{-1})$, then
$\alpha(t)= \mathfrak{exp}(tv) \circ \eta$ is the unique geodesic
joining $\eta$ to $\varphi$ in $\mathcal{U}(\eta)$ cf. Theorem \ref{theorem3}.

To prove the second statement, let
$\varphi\circ\eta^{-1}=\mathfrak{exp}(r\,w)$ with $\Vert w\Vert_k
=1$ and choose $\varepsilon \in (0,r)$. If $\gamma$ is any
piecewise $C^1$-curve on $\Diff$ joining $\eta$ to $\varphi$, then
$\gamma$ contains an arc of curve $\gamma^\ast$ such that, after
reparametrization,
\begin{equation*}
\norm{ \mathfrak{exp}^{-1}(\gamma^\ast(0)) }_k= \varepsilon,\quad
\norm{ \mathfrak{exp}^{-1}(\gamma^\ast(1)) }_k=r,
\end{equation*}
and
\begin{equation*}
\varepsilon \le \norm{ \mathfrak{exp}^{-1}(\gamma^\ast(t)) }_k \le
r,\quad t \in [0,1].
\end{equation*}

Lemma \ref{lemma3} yields $l(\gamma^\ast) \ge r-\varepsilon$, thus
$l(\gamma) \ge l(\gamma^\ast)\ge r- \varepsilon$. The
arbitrariness of $\varepsilon>0$ ensures $l(\gamma) \ge r$. But
$l(\alpha)=r$ in view of Lemma \ref{lemma3} and the minimum is
attained if and only if the curve is a reparametrization of a
geodesic.
\end{proof}


\section{Comments}

This section is devoted to a discussion of the $L^2( \Circle)$
right-invariant metric on $\Diff$, case when the geodesic equation
\eqref{EulerEquation} is the inviscid Burgers equation
\begin{equation*}
u_t+3uu_x=0
\end{equation*}
cf. \cite{ArnKhe}. The crucial difference from the case of a
$H^k(\Circle)$ right-invariant metric (with $k \ge 1$) lies in the
fact that the inverse of the operator $A_k$, defined by
\eqref{InertiaOperator}, is not regularizing. This feature makes
the previous approach inapplicable but the existence of geodesics
can be proved by the method of characteristics.\bigskip

\begin{prop}[\cite{ConKol}]\label{proposition1}
For the $L^2( \Circle)$ right-invariant metric on $\Diff$ there
exists a unique smooth geodesic on $\Diff$ starting at $Id$ in the
direction $u_0 \in T_{Id}\Diff$.
\end{prop}

This result enables one to define the Riemannian exponential map
of the $L^2( \Circle)$ right-invariant metric on $\Diff$, in
analogy to the cases considered in the present paper.
However,\bigskip

\begin{prop}[\cite{ConKol}]\label{proposition2}
The Riemannian exponential map of the $L^2( \Circle)$
right-invariant metric on $\Diff$ is not a $C^1$-diffeomorphism
from a neighborhood of zero in $T_{Id}\Diff \equiv
C^\infty(\Circle)$ to a neighborhood of the identity on $\Diff$.
\end{prop}

The question whether another right-invariant metric could provide
$\Diff$ with a nice Riemannian structure has been positively
answered in this paper.


\bibliographystyle{amsplain}
\bibliography{hk}
\end{document}